\documentclass[conference]{IEEEtran}
\IEEEoverridecommandlockouts
\usepackage{cite}
\usepackage{amsmath,amssymb,amsfonts}
\usepackage{algorithmic}
\usepackage{graphicx}
\usepackage{textcomp}
\usepackage{xcolor}
\usepackage{physics}
\usepackage{caption}
\usepackage{subcaption}

\usepackage{hyperref}

\usepackage{tikz}
\usetikzlibrary{quantikz2}

\def\BibTeX{{\rm B\kern-.05em{\sc i\kern-.025em b}\kern-.08em
    T\kern-.1667em\lower.7ex\hbox{E}\kern-.125emX}}

\makeatletter 
\newcommand{\linebreakand}{%
  \end{@IEEEauthorhalign}
  \hfill\mbox{}\par
  \mbox{}\hfill\begin{@IEEEauthorhalign}
}
\makeatother 

\begin{document}

\title{Hamiltonian-based Quantum Reinforcement Learning for Neural Combinatorial Optimization}

\author{
\IEEEauthorblockN{Georg Kruse}
\IEEEauthorblockA{\textit{Fraunhofer IISB} \\
\textit{Technical University Munich} \\
Erlangen, Germany \\
georg.kruse@iisb.fraunhofer.de}
\and
\IEEEauthorblockN{Rodrigo Coelho}
\IEEEauthorblockA{\textit{Fraunhofer IISB} \\
Erlangen, Germany \\
rodrigo.coelho@iisb.fraunhofer.de}
\and
\IEEEauthorblockN{Andreas Rosskopf}
\IEEEauthorblockA{\textit{Fraunhofer IISB} \\
Erlangen, Germany \\
andreas.rosskopf@iisb.fraunhofer.de}
\and
\linebreakand
\IEEEauthorblockN{Robert Wille}
\IEEEauthorblockA{\textit{Technical University Munich} \\
Munich, Germany \\
robert.wille@tum.de}
\and
\IEEEauthorblockN{Jeanette Miriam Lorenz}
\IEEEauthorblockA{\textit{Fraunhofer IKS} \\
\textit{Ludwig Maximilian University} \\
Munich, Germany \\
jeanette.miriam.lorenz@iks.fraunhofer.de}
}
\maketitle

\begin{abstract}

Advancements in Quantum Computing (QC) and Neural Combinatorial Optimization (NCO) represent promising steps in tackling complex computational challenges. On the one hand, Variational Quantum Algorithms such as QAOA can be used to solve a wide range of combinatorial optimization problems. On the other hand, the same class of problems can be solved by NCO, a method that has shown promising results, particularly since the introduction of Graph Neural Networks. Given recent advances in both research areas, we introduce Hamiltonian-based Quantum Reinforcement Learning (QRL), an approach at the intersection of QC and NCO. We model our ansatzes directly on the combinatorial optimization problem's Hamiltonian formulation, which allows us to apply our approach to a broad class of problems. Our ansatzes show favourable trainability properties when compared to the hardware efficient ansatzes, while also not being limited to graph-based problems, unlike previous works. In this work, we evaluate the performance of Hamiltonian-based QRL on a diverse set of combinatorial optimization problems to demonstrate the broad applicability of our approach and compare it to QAOA.

\end{abstract}


\section{Introduction}

Recently, both Quantum Computing (QC) and Neural Combinatorial Optimization (NCO) have seen significant progress in their respective areas, both offering (potential) solutions to complex computational problems \cite{preskill2018quantum, bello2016neural}. On the one hand, in NCO, Reinforcement Learning (RL) algorithms can be used  to learn heuristics to find approximate solutions to Combinatorial Optimization (CO) problems \cite{bello2016neural}. This is particularly relevant considering many CO problems are NP-hard due to their exponential scaling of solution space. On the other hand, QC is believed to solve problem classes that are intractable for classical computers. Moreover, Variational Quantum Algorithms (VQAs) for tackling CO problems, such as QAOA \cite{farhi2014quantum}, already exist on currently available quantum hardware. QAOA has been extensively studied in this context, particularly how to design cost and mixer Hamiltonians and its effect on solution quality \cite{brandhofer2022benchmarking}. Nevertheless, quantum advantage still remains to be demonstrated.

Given the recent progress in these two research areas, their combination might be particularly suited to solve complex CO problems. In this context, Quantum Reinforcement Learning (QRL) can be used to learn heuristics following the paradigm of classical NCO. In QRL, the neural network of the classical RL algorithm is replaced by a Variational Quantum Circuit (VQC) as a function approximator, which is then optimized by a classical training loop. While the ansatzes of VQAs, such as QAOA, are problem dependent, in QRL, the problem-agnostic hardware efficient ansatz (HEA) in different shapes and forms is widely used \cite{druagan2022quantum, kruse2023variational}. However, the main advantage of the HEA might as well be its main disadvantage: being problem-agnostic. While it can be applied to a large class of problems due to its high expressibility and its problem independence, it becomes untrainable due to Barren Plateaus (BPs) \cite{mcclean2018barren, holmes2022connecting} as problem sizes increase. Hence, most works in QRL focus on small-sized toy-problems with less than a dozen qubits \cite{skolik2022quantum, coelho2024vqc}, with their scalability to larger problem instances being highly unlikely.

Therefore, recent works have focused on VQC architectures with favourable properties that make them less prone to BPs.  One line of research shows that an increase in parameter correlations within a given ansatz improves its trainability \cite{holmes2022connecting}. Similarly, Larocca et al. argue that ansatzes can only be effectively trained if their generator sets do not yield full-rank Dynamical Lie Algebras (DLA) \cite{larocca2022diagnosing}, and Ragone et al. introduce a general theory of BPs, illustrating various strategies on how to avoid them \cite{ragone2023unified}. Another line of research has been explored by Skolik et al. \cite{skolik2023equivariant} and Mernyei et al. \cite{mernyei2022equivariant}, who argue that symmetry-preserving quantum circuits show better training performance than their non symmetric counter parts. Additionally, Skolik et al. propose a first step towards the integration of QC into NCO. They introduce the Q-Learning algorithm for learning on weighted graphs and showcase that equivariant quantum circuits outperform standard HEAs on the traveling salesperson problem (TSP). Furthermore, He \cite{he2024quantum} has proposed to combine classical graph neural networks and quantum annealing to solve the TSP.

Based on the considerations on the trainability of ansatzes and the initial integration of QC into NCO, our contributions are as follows: We introduce Hamiltonian-based QRL, where we base our ansatz directly on the (problem) Hamiltonian of (binary) CO problems (Section \ref{H_based_QRL}). We analyse the trainability of our ansatz (Section \ref{sec:trainability}) and show its relation to QAOA (Section \ref{sec:relation_qaoa}). In order to reduce the number of additionally required qubits due to inequality constraints, we incorporate an encoding strategy by \cite{montanez2022unbalanced} in our approach (Section \ref{sec:inequality_encoding}). The broad applicability of our ansatz is demonstrated on three CO problems (Section \ref{sec:cop}). Finally, we compare our approach to previously introduced QRL methods and analyse its training performance with a detailed comparison against QAOA (Section \ref{sec:numerics}).

\begin{figure}[ht] 
\includegraphics[width=\linewidth]{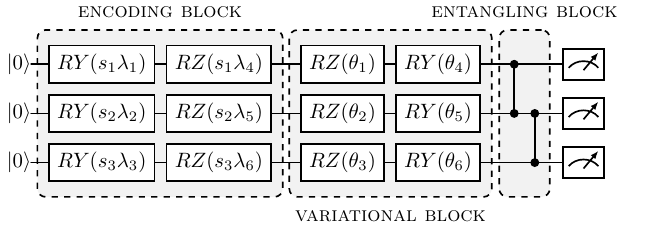}
\caption{Single-layer VQC for QRL: $U(s,\theta)$ generally consists of three blocks which are repeated in each layer: An \emph{encoding-block}, where the features of state $s$ (possibly scaled by additional trainable parameters $\lambda$) are encoded. A \emph{variational-block}, where additional parameterized quantum gates are placed, and an \emph{entangling-block}, where the entanglement gates are placed. However, this structure is not static and blocks may be switched or merged with one another.}\label{fig:vqc_hea}
\end{figure}

\section{Quantum Reinforcement Learning}\label{QRL}

An RL problem consists of two parts: the agent (the decision maker consisting of a function approximator that is trained) and the environment (the surroundings with which the agent interacts). The goal of the agent is to learn how to interact with the environment such that the reward signal it receives from this interaction is maximized. Most RL environments are based on a \emph{Markov Decision Process} (MDP). In short, it consists of a set of states $\mathcal{S}$, a set of actions $\mathcal{A}$, a state transition probability function $\mathcal{P}$ (that defines the probability of transitioning from a state $s$ to a state $s\prime$ after taking action $a$) and a reward function $\mathcal{R}$, that guides the RL agent's training. 

The behavior of the agent is represented by the policy $\pi(s,a)$, which is a probability function of actions conditioned on states. The policy is updated such that it maximizes the cumulative return $G_t$, defined as

\begin{equation}
    G_t = \sum_{k=0}^{\infty}\gamma^{k}R_{t+k+1},
\end{equation}

where $\gamma \in[0,1]$ is a discount factor and $R_t$ the reward at time step $t$. In any MDP, there exists at least one deterministic optimal policy $\pi^*(s,a)$ that maximizes the cumulative return $G_t$. Hence, the goal of the RL agent is to learn this policy. In classical RL, the most common approach to approximate the optimal policy $\pi^*(s,a)$ is by using a neural network as function approximator for the policy $\pi(s,a)$. In the subfield of QRL which this work is confined to, the sole modification is the replacement of the neural network with a VQC, with the (classical) RL algorithms themselves remaining (except for the hyperparameters) unchanged. Hence, the main difference stems from the function approximator which is either classical or quantum. A typical VQC $U(s,\theta)$ that can be used as function approximator for QRL takes the classical environment state $s$ as input and is parameterized by the trainable parameters $\theta$. The general structure of a VQC for QRL based on the HEA is depicted in Fig. \ref{fig:vqc_hea}. In the following sections, we briefly cover the two RL algorithms used throughout this work and their adaptations to QRL. For a broader explanation of the classical RL algorithms, the reader is referred to \cite{sutton2018reinforcement}.

\subsection{Q-Learning}

In Q-Learning, the \emph{action-value function} $q_\pi(s_t,a_t)$ defines the expected value of taking action $a_t$ in state $s_t$ as

\begin{equation}
    q_\pi(s,a) = \mathbb{E}_\pi\left[G_t|s_t=s,a_t=a\right]
\end{equation}

where $G_t$ is the cumulative return at time step $t$. In state $s_t$ the next action is chosen according to $a_t = argmax \, Q(s_t, a)$. To ensure initial exploration, a so called $\epsilon$-greedy policy can be applied such that the probability of choosing a random action is $1-\epsilon$. Furthermore, $\epsilon$ typically decays over time to balance the exploration-exploitation trade-off \cite{sutton2018reinforcement}. In QRL, the Q-value of a given state action pair $Q(s,a)$ is computed by the expectation value of an observable $O_a$

\begin{equation}
    Q(s,a) = \bra{0^{\otimes n}}U(s,\theta)^{\dagger}O_aU(s,\theta)\ket{0^{\otimes n}}
\end{equation}

where $n$ is the number of qubits and $U(s,\theta)$ is a VQC that depends on the state $s$ and parameters $\theta$. The parameters are then updated using a variant of the stochastic gradient descent algorithm on

\begin{equation}
[r+\gamma\max_{a\prime}\hat{Q}(s\prime,a\prime,\theta_{i}^{-} - Q(s,a,\theta_i))^2]
\end{equation}

where $\hat{Q}$ is the \emph{target network}, an additional VQC $\hat{U}(s,\hat{\theta})$ with frozen weights that are updated every $C$ time steps to ensure training stability. In the following sections, we refer to the Q-learning algorithm with a trainable VQC at its core as QDQN.

\subsection{Policy Gradient}

The Policy Gradient (PG) algorithm aims to find a parameterized policy $\pi_{\theta}$ which maximises the cumulative reward in a given environment. At each time step $t$, the agent chooses an action $a_t$ in a given state $s_t$ with probability $\pi_{\theta}(a_t|s_t)$, such that the cumulative return $G_t$ is maximized. In QRL, the trainable parameters $\theta$ of the function approximator $U(s, \theta)$ are optimized according to the gradient of the performance measure $J(\theta)$

\begin{equation}
    \nabla_{\theta} J(\theta) = \mathbb{E}_{\pi_{\theta}} \Big[ \sum \nabla_{\theta} ln ( \pi_{\theta} (a_t | s_t) )\cdot G_t\Big].
\end{equation}

For a given VQC $U(s, \theta)$ acting on $n$ qubits, we can compute the probability of an action $a_i$ according to the current policy $\pi_{\theta}$ by measuring an observable $O_{a_i}$. As has been shown in previous works, the choice of observables can have a significant impact on the training performance \cite{meyer2023quantum}. In this work we introduce a slight modification to the approach originally proposed by \cite{jerbi2021parametrized}: To improve the training performance of the VQC, we add an additional trainable parameter that scales the expectation values of the observables, an approach similar to the one introduced in the context of Q-Learning \cite{skolik2022quantum} or actor-critic algorithms \cite{kruse2023variational}. In the following we refer to the PG algorithm with a trainable VQC at its core as QPG.

\subsection{Quantum Reinforcement Learning for Neural Combinatorial Optimization}\label{QRL_for_NCO}

The general idea behind QRL for NCO is to train a QRL model to learn a heuristic capable of solving CO problems. Moreover, we want our models to be able to generalize fairly well to unseen problem instances. That way, one trains the model once on a dataset of problem instances and then uses its heuristics to find solutions for other new and unseen problem instances. Other classical optimization algorithms such as branch-and-bound algorithms or VQAs such as QAOA and VQE need to be run on each new problem instance individually, hence the computational cost for each new problem instance remains the same. In NCO, the computational cost of the algorithm is instead moved to a previous training phase. While this leads to higher computational costs during this phase, it enables fast predictions for new unseen problem instances with only minor additional computational costs after training. A trained NCO model can hence outperform other algorithms in terms of computational cost and, therefore, on runtime for new and unseen problem instances \cite{bello2016neural}.

Even though CO problems can contain time dependent variables, their description as a graph or QUBO is static. One way to solve such a CO with an RL algorithm is to model it as a \textit{multiarmed bandit} problem: A \textit{bandit} problem can be thought of as a special MDP, where each episode consists of a single time step. \textit{Multiarmed} in that context means that multiple actions can be chosen in one time step. Moreover, the state of the environment may vary between episodes, giving rise to a \textit{contextual multiarmed bandit}. In this setting, the RL agent learns on a dataset of problems, where each episode consists of one single randomly sampled problem instance. All (binary) decision variables are chosen in a single time step. Hence, the action space in this formulation is multi-discrete: For all (binary) decision variables of the CO problem instance, the agent assigns either a "0" or a "1" in a single time step. The output (action) is a vector of size corresponding to the number of decision variables of the problem instance. In QRL, this can be realized by encoding the static problem instance as state $s$ into a unitary $U(s, \theta)$ and conducting a measurement with observables $O_{a_i}$ such that all variable values of the problem instance can be derived (ref. Section \ref{sec:cop}). 

Another way to solve CO problems with an RL algorithm is to assign one decision variable at each time step, constructing the solution to the problem one variable at a time, similar to the approach used in  \cite{skolik2023equivariant} and \cite{khalil2017learning}. To do this, additional information needs to be added to the static problem information, namely annotations \cite{skolik2023equivariant}. For a given QRL agent with unitary $U(s, \alpha, \theta)$ as function approximator, the state $s$ encodes the static problem information (e.g. the Hamiltonian of the sampled instance that remains static throughout the episode), while the annotations $\alpha$ encode the current status of the episode (e.g. which variable has been previously assigned and which are free to be assigned). Using this method, the QRL agent can distinguish between different time steps in the same episode and sequentially build the final solution with appropriate measurements (ref. Section \ref{sec:cop}). \\

Because ansatzes such as the HEA  have been shown to be untrainable due the BP phenomena, \cite{skolik2023equivariant} and  \cite{mernyei2022equivariant} propose to use symmetry-preserving quantum circuits that perform better than their non symmetric counter parts. Skolik et al. are the first to show in a QRL for NCO setting how to construct an ansatz from a weighted graph which is equivariant under node permutations. For this problem-inspired encoding scheme, the edges of a graph (corresponding to the weights of the weighted-MaxCut or the distances of the TSP) are encoded as two-qubit gates. The use of only a single layerwise parameter for each \emph{encoding-block} and a single layerwise parameter for each \emph{annotation-block} (which can be seen as an in-between of the \emph{encoding-block} and the \emph{variational-block}) ensures that the ansatz remains equivariant throughout training. They prove that the Q-Learning algorithm for learning on weighted graphs preserves the equivariance property as long as no additional individual output scaling parameters are introduced. 

\section{Hamiltonian-based Quantum Reinforcement Learning}\label{H_based_QRL}

To solve CO problems with QRL, we base our ansatz on the (problem) Hamiltonian of the given binary CO problem directly. Instead of encoding the graph information of the problem, as has been introduced by \cite{skolik2023equivariant}, we take the QUBO formulation of the problem and use the structure of its Hamiltonian representation as ansatz for our QRL agent. Hence, we can expand the approach to the same range of problems as QAOA instead of being limited to (weighted) graph problems. Any binary CO problem that can be mapped to a QUBO formulation can hence also be solved by Hamiltonian-based QRL. However, even though the structure of the ansatz is directly based on the QUBO formulation, it remains an open question how the ansatz should be parameterized and whether or not additional trainable gates should be added to the ansatz. To answer this question, one first needs to consider the effects parameterization and additional gates have on the trainablity of the ansatz, which we will analyse next. 

The trainabilty of ansatzes has been subject of extensive study  \cite{schatzki2024theoretical, larocca2023theory, wiersema2309classification}. In order to successfully train a VQC, the chosen ansatz must not only contain the desired solution, but also exhibit large enough cost function gradients \cite{holmes2022connecting}, such that gradient based optimization is feasible. In \cite{ragone2023unified} and \cite{diaz2023showcasing}, four main sources for BPs have been identified: An excess in circuit expressivity, generalized entanglement, global measurements and circuit noise. In order to theoretically assess whether or not an ansatz exhibits BPs, one can analyse the sets of generators the ansatz of interest is constructed from. In \cite{larocca2022diagnosing}, the dimension of the DLA is introduced as a measure for the rate at which the variance of the gradient of an ansatz vanishes. As an example, Larocca et al. show that the dimension of the DLA of the generator sets of the  Erdös-Rényi model - better known as the MaxCut Hamiltonian - as well as the widely spread Spin Glass model is of full rank and hence the variance of the cost function vanishes exponentially with system size. However, even though both ansatzes exhibit BPs, the rate at which the variance of the gradients vanishes differs and depends on the amount of layers as well as on their respective generator sets. 

Following the same principle, but from an empirical point of view, Holmes et al. have also found the excess in circuit expressivity to be a cause of BPs \cite{holmes2022connecting}. To reduce the expressivity and thereby increase the trainability of an ansatz, they propose to correlate the trainable parameters. By utilizing the same parameters across qubits in each layer (as it is generally done for QAOA where the $\beta$ and $\gamma$ parameters are used across all respective Hamiltonian gates within one layer), they are able to reduce the slope of the variance decay of the gradients. According to \cite{holmes2022connecting}, the most effective strategies to amplify gradients are the correlation of parameters and the reduction of circuit depth.

\begin{figure}[ht!]
    \centering
    \includegraphics[width=\linewidth]{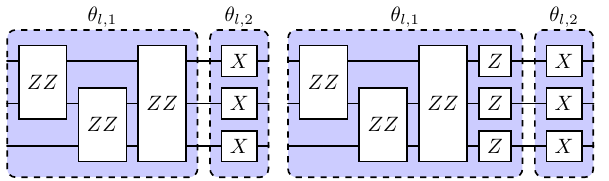}
    \caption{Schematic illustration of a single layer of a three qubit VQC for the sets of generators $G_{PP,P}$ in Eq. \ref{generator_1} (left) and $G_{PP+P,P}$ in Eq. \ref{generator_2} (right) of the \emph{sge-sgv} ansatz.}
    \label{fig:sge-sgv}
\end{figure}

\begin{figure}[ht!]
    \centering
    \includegraphics[width=0.9\linewidth]{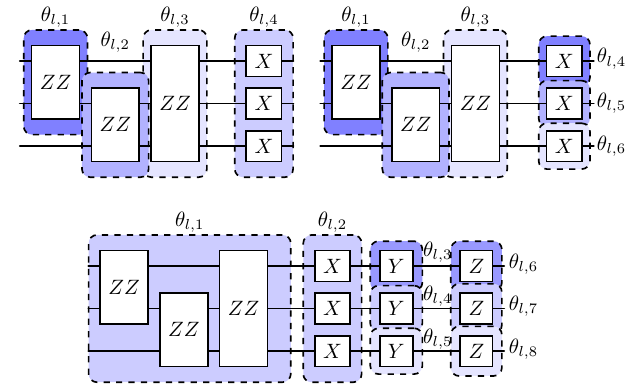}
    \caption{Schematic illustration of a single layer of a three qubit VQC for the sets of generators $G_{mge-sgv}$ in Eq. \ref{eq:mge-sgv} (upper left), $G_{mge-sgv}$ in Eq. \ref{eq:mge-mgv} (upper right), and $G_{sge-sgv+hea}$ in Eq. \ref{eq:sge-sgv+hea} (lower center).}
    \label{fig:mge}
\end{figure}

Based on these considerations, we propose to use an ansatz constructed from a minimal set of generators, using as much correlation as possible: Our ansatzes for Hamiltonian-based QRL are constructed using the set of generators $G_{PP,P}$, if the QUBO formulation contains only quadratic terms, and the set $G_{PP+P, P}$, if the QUBO formulation has additional linear terms, with $P$ being an arbitrary Pauli gate $P \in {X, Y, Z}$ and $PP$  two-qubit Pauli gates $PP \in {XX, YY, ZZ}$. 

\begin{equation} \label{generator_1}
    G_{PP,P} = \Big\{\sum^n_{i<j} P_i P_j, \sum^n_{i=1} P_i \Big\}.
\end{equation}

\begin{equation}\label{generator_2}
    G_{PP+P, P} = \Big\{\sum^n_{i<j} P_i P_j + \sum^n_{i=1} P_i, \sum^n_{i=1} P_i \Big\}
\end{equation}

\begin{figure}[ht] 
\centering
\includegraphics[width=0.9\linewidth]{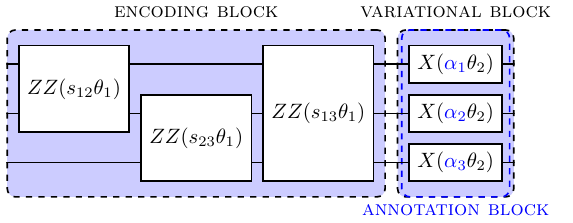}
\caption{Hamiltonian-based QRL: A single layer of the \emph{sge-sgv} ansatz consists of an \emph{encoding-block}, where the features of state $s$ with trainable parameters $\theta$ are encoded, and a \emph{variational-block}, where additional parameterized quantum gates are placed. If additional annotations $\alpha$ are used, we also refer to the \emph{variational-block} as \emph{annotation-block}.}\label{fig:qrl}
\end{figure}

In Eq. \ref{generator_1} and \ref{generator_2}, we use a single generator and hence a single parameter for our \emph{encoding-block} and one additional generator as \emph{variational-block} (or \emph{annotation-block} ref. Fig. \ref{fig:qrl}), totalling two parameters per layer. We call this ansatz in the following  \emph{sge-sgv} (\emph{single generator encoding-block - single generator variational-block}). The ansatzes are depicted in Fig. \ref{fig:sge-sgv}.

Additionally, we also propose variants of the sets $G_{PP,P}$ and $G_{PP+P,P}$. Starting with \emph{sge-sgv}, we introduce two modifications to this ansatz: \emph{multiple generator encoding-block - single generator variational-block} (\emph{mge-sgv}) and \emph{multiple generator encoding-block - multiple generator variational-block}  (\emph{mge-mgv}). Hence, the generator sets for CO problems with QUBO formulations without linear terms are 

\begin{equation} \label{eq:mge-sgv}
    G_{mge-sgv} = \Big\{ P_i P_j \Big\}^n_{i<j} \bigcup \Big\{\sum^n_{i=1} P_i\Big\}
\end{equation}

for the \emph{mge-sgv} ansatz and

\begin{equation} \label{eq:mge-mgv}
    G_{mge-mgv} = \Big\{  P_i P_j \Big\}^n_{i<j} \bigcup \Big\{  P_i \Big\}^n_{i=1}
\end{equation}

for the \emph{mge-mgv} ansatz respectively (ref. Fig. \ref{fig:mge}). While the amount of entanglement of these ansatzes remains unchanged, the increase of generators and therefore the decrease of parameter correlation is expected to lead to a faster decrease of gradient variance (ref. Section \ref{sec:trainability}). Additionally, we compare our ansatzes to an \emph{sge-sgv} ansatz, to which we add an additional HEA (ref. Fig. \ref{fig:mge}). The generator set of the ansatz is hence

\begin{equation}\label{eq:sge-sgv+hea}
    G_{sge-sgv+hea} =  G_{sge-sgv} \bigcup \Big\{Y_i, Z_i  \Big\}^n_{i=1}.
\end{equation}

In summary, constructing our ansatzes solely from $G_{PP,P}$ and $G_{PP+P,P}$ for CO problems gives us various advantages: 
First, we encode a meaningful representation of the CO problem as our ansatz. It incorporates the structure of the problem directly, which can greatly improve training performance as has been previously shown for graph neural networks \cite{schuetz2022combinatorial}. 
Second, we greatly increase the trainability of our ansatz by strongly correlating the parameters within one layer \cite{holmes2022connecting}. 
Third, we preserve the equivariance property  of our ansatz during training \cite{skolik2023equivariant} (if no additional individual output scaling is applied), which can improve the performance of our ansatz, especially on equivariant problems. 
Fourth, the low number of parameters compared to other ansatzes accelerates training, mainly on real quantum hardware, where at least two circuit executions are needed to estimate the gradient of a parameter with the parameter shift rule \cite{wierichs2022general}. 

In the following, we analyse the trainability of our ansatz and  illustrate our approach on exemplary CO problems in Section \ref{sec:cop}. 

\subsection{Trainablity of Ansatz} \label{sec:trainability}

\begin{figure}
    \centering
    \includegraphics[width=0.8\linewidth]{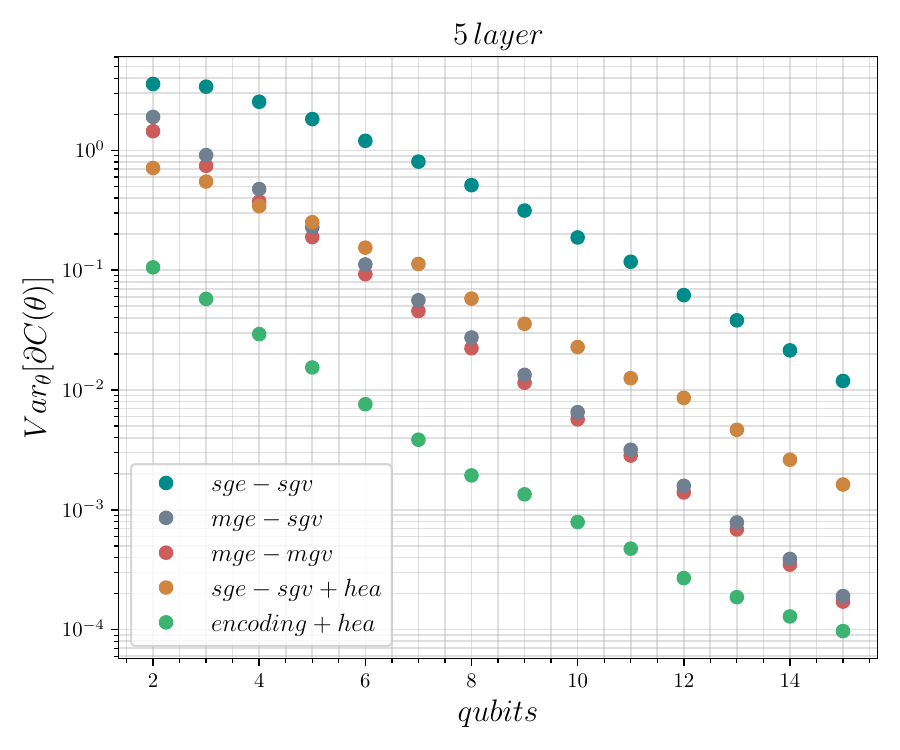}
    \caption{Numerical results of the variance of the cost function partial derivatives for the introduced ansatzes with $L=5$ layers. For each point, we evaluated the variance of the gradient for 1000 samples.}
    \label{fig:variance}
\end{figure}

The generator sets $G_{PP,P}$ and $G_{PP+P,P}$ of the proposed \emph{sge-sgv} ansatz for QRL are closely related to the Erdös-Rényi model and the Spin Glass model, respectively. Larocca et al. show that ansatzes which are built from these generator sets also have BPs \cite{larocca2022diagnosing} (for circuit depths at which the ansatzes approximate a 2-design). Larocca et al. analyse the variance of the gradients of the generator set of the Erdös-Rényi model with $L=12n$ layers (with $n$ being the number of qubits) and of the generator set of the HEA with $L=100$ layers. However, ansatzes with hundreds of layers are out of the scope of current hardware and impractical for QRL due to their higher numbers of parameters. We therefore investigate the gradient scaling for shallow circuits of $L = 5$ layers for all previously introduced ansatzes (with linear terms) and compare it to an ansatz with no trainable parameters in the \emph{encoding-block} and an HEA ansatz as \emph{variational-block} (\emph{encoding + hea}). In all cases we take the partial derivative with respect to the parameter $\theta_{\frac{L}{2},2}$ as proposed by \cite{larocca2022diagnosing}. For each ansatz, we compute the variance of the gradient by sampling 1000 random initializations. In Fig. \ref{fig:variance} it can be seen that the variance of the gradient vanishes exponentially for all evaluated ansatzes. However, the offset as well as the rate at which the variance vanishes differs: The \emph{mge-sgv} and \emph{mge-mgv} ansatzes' variance decays quickly, with a slightly higher offset of the \emph{mge-sgv} due to the increased parameter correlation in the variational block. The \emph{sge-sgv} ansatz exhibits the best variance scaling. The variance of the \emph{sge-sgv+hea} and the \emph{encoding+hea} ansatzes decay at a similar rate, but with a lower offset, and at $10^{-4}$ the decay of the \emph{encoding+hea} ansatz starts to decelerate due to the small amount of used layers \cite{mcclean2018barren}. This shows that the proposed \emph{sge-sgv} ansatz shows the highest gradient variance and therefore the best trainability for the problem sizes used in this work.

\subsection{Solving Combinatorial Optimization Problems with Hamiltonian-based Quantum Reinforcement Learning}\label{sec:solving_cop}

In Hamiltonian-based QRL, each binary decision variable of the QUBO corresponds to a single qubit. Thus, for a given problem instance with $n$ binary decision variables (and no additional constraints), our encoding  requires $n$ qubits. During the training of the QRL agent, at the beginning of each episode a problem instance is sampled randomly from a training dataset and, at the end of the episode, the QRL agent returns a binary vector with the final value for each of the decision variables. For example, if for a given MaxCut instance with three nodes $a,b,c$, the QRL agent outputs the final action $[0,1,1]$, it means that the node $a$ was given value $0$ and so on. The way the agent outputs the binary variable vector depends on the problem being solved and its formulation as an RL environment. It is either done sequentially, where for each time step, the RL agent assigns one of the variables, constructing the solution one variable at a time, or it is solving each instance as a bandit problem, assigning all decision variables in a single time step.

\subsubsection{Relation to QAOA}\label{sec:relation_qaoa}

A popular approach to solve CO problems is to use QAOA \cite{gleissner2023restricted, palackal2023quantum}, which finds an approximate solution to the given problem. For QAOA to be used, the problem instance first needs to be reformulated as a QUBO, which has a trivial translation to the Ising model (the problem Hamiltonian). However, most CO problems of interest have constraints, which need to be encoded as penalties in the QUBO formulation. While the encoding of equality constraints is trivial \cite{glover2019quantum}, this can be quite complex for inequalities (ref. Section \ref{sec:inequality_encoding}). Once the QUBO formulation is constructed, QAOA can be used to approximate the ground state of the corresponding problem Hamiltonian. In this context, QAOA can be seen as a variant of the Hamiltonian-based QRL method for solving a bandit problem where all decision variables are assigned in a single time step via a single global observable which acts on all qubits. On the contrary, in the proposed QRL bandit approach, a single-qubit observable is applied for each decision variable.

\subsubsection{Inequality Encoding}\label{sec:inequality_encoding}

There are mainly two approaches for encoding inequality constraints as penalties: \emph{unbalanced penalization} \cite{montanez2022unbalanced} and \emph{slack variables} \cite{vyskovcil2019embedding}. The later consists of turning the inequality constraints into equality constraints by the use of \emph{slack variables}, at the expense of additional qubits. The former consists of approximating the constraints using (a quadratic approximation of) the exponential decay function (ref. Eq. \ref{eq:exponential}), effectively yielding an exponential penalty if the constraint is broken and a negligible penalty if not. 

\begin{equation}\label{eq:exponential}
    e^{-h(x)} \approx 1 - h(x) + \frac{1}{2} h(x)^2
\end{equation}

Montanez et al. show that the \emph{unbalanced penalization} method outperforms the \emph{slack variables} approach in two ways: It does not require additional qubits and the optimal solution is found with higher probability \cite{montanez2022unbalanced}. However, this method has one major drawback, as there is no guarantee that the ground state of the encoded Hamiltonian corresponds to the optimal solution of the original problem. Even though the authors empirically show that the optimal solution is typically among the lowest eigenvalues of the Hamiltonian, this still means that QAOA might be optimizing for the wrong solution, possibly even an invalid one. 

By adapting the \emph{unbalanced penalization} method to Hamiltonian-based QRL, we are able to maintain the main advantage of the approach (the reduced number of required qubits) while coping with its main disadvantage (potentially not optimizing towards the optimal solution). We start with the \emph{unbalanced penalization} method and reformulate the inequalities of the CO problem as a QUBO without the need for ancilla qubits. This QUBO is then mapped to the problem Hamiltonian and used as the ansatz for the VQC of our QRL agent. However, we choose a reward function that depends only on the original problem, not on the ground state of the problem Hamiltonian. This way we can ensure that, even if the Hamiltonian's ground state does not encode the optimal solution of the CO problem, the reward function of our QRL agent is still an unbiased guide towards the optimal solution. Additionally, we drive the QRL agents towards valid solutions,  by either masking solutions that would break the constraints (hard-constraint) or simply penalizing the agent if it outputs an action that does so (soft constraint).

In summary, when using QAOA, one is limited to minimizing a Hamiltonian that might not correctly encode the original problem, leading possibly to an incorrect or even invalid solution. Hamiltonian-based QRL, on the other hand, combines the best of both worlds. It uses the \emph{unbalanced penalization} Hamiltonian as ansatz, effectively incorporating information about the problem with less required qubits compared to the standard \emph{slack variables} approach. However, using a reward function that represents the ground truth of the original problem for training mitigates the main disadvantage of the \emph{unbalanced penalization} method when applied to QAOA.  \\

\subsubsection{Combinatorial Optimization Problems}\label{sec:cop}

To demonstrate the applicability of Hamiltonian-based QRL to a wide class of binary CO problems, we evaluate our agents on three diverse problems and showcase different variants of our approach on the weighted-MaxCut, the Unit Commitment Problem (UCP) and the Knapsack Problem (KP). In the following, we formulate these problems as RL environments and show how to use Hamiltonian-based QRL to solve them.

\paragraph{Weighted-MaxCut}\label{sec:maxcut_env}
For a given graph $G=(V,E)$ with nodes $V$ and edges $E$ of size $n$ and edge weights $w$, the objective is to partition its nodes into two disjoint sets such that the total weight of the edges connecting the two sets is maximized. 

\begin{equation}\label{eq:maxcut}
    max_{x\in\{0,1\}} \sum_{ij \in E} w_{i,j} (1 -  x_{i}x_{j})
\end{equation}

In Eq. \ref{eq:maxcut}, $x_i$ is the binary variable representing whether vertex $i$ is in set 0 or set 1. To solve this problem using Hamiltonian-based QRL, the set of generators as well as the observables need to be defined: Since the problem statement of Eq. \ref{eq:maxcut} does not contain linear terms, we chose the generator set for the \emph{sge-sgv} ansatz to be

\begin{equation} \label{eq:g_maxcut}
    G_{maxcut} = \Big\{\sum^n_{i<j} Z_iZ_j, \sum^n_{i=1} X_i \Big\},
\end{equation}

using the first generator to encode the static problem and the second generator for annotation (ref. Section \ref{QRL_for_NCO} and Fig. \ref{fig:qrl}). In our RL environment's formulation of the weighted-MaxCut (mainly following the approach of \cite{khalil2017learning}), a single episode of the game starts with a randomly sampled graph instance from a dataset with all nodes assigned to a single set (set 0), and with all annotation terms $\alpha$ set to $\pi$. At each time step, the QRL agent assigns one node to the second of the two sets (set 1). The action is selected either by node measurements using $X$ observables (following our approach), or by edge measurements (as proposed by \cite{skolik2023equivariant}) using $ZZ$ observables without additional output scaling. After each action, the change between the cut weight of the previous and the current cut is returned as reward.  After a node has been assigned to set 1 at time step $t$, it is masked as a possible action for the following time steps, such that the agent cannot select the same node several times during an episode. All previously assigned nodes are annotated for the next time step $t+1$ and their respective value of $\alpha$ is set to 0. The episode ends as soon as the change in cut weight is either 0 or negative. The problem instances used in our experiments are taken from \cite{skolik2023equivariant}.

\paragraph{Unit Commitment}\label{sec:uc_env}

For a given set of $N$ power generators with binary operating status $x$, output power $p$, minimal and maximal output power $p_{min}$ and $p_{max}$, the goal of the UCP is to satisfy a power demand $L$ such that the cost of energy generation is minimized.

\begin{equation}\label{eq:uc_obj}
    min_{x\in\{0,1\}} \sum^{N}_{i=1} (A_i + B_ip_i + C_ip_i^2)x_i
\end{equation}

\begin{equation}\label{eq:uc_const}
    \sum^{N}_{i=1} p_ix_i = L
\end{equation}

In the objective function Eq. \ref{eq:uc_obj}, the factors $A$, $B$ and $C$ represent the cost of power generation of the respective generator and the values are taken from the dataset of \cite{kazarlis1996genetic}. In the simplest scenario of the UCP, the objective function Eq. \ref{eq:uc_obj} is subject only to the equality constraint of Eq. \ref{eq:uc_const} \cite{koretsky2021adapting}. For the standard version of the UCP, the classical solver needs to assign binary values to $x$ and integer values to $p$. In more realistic UCP scenarios, these values need not only to be assigned for a single time step, but rather a sequence of time steps with additional time dependent constraints for generator up and down times (see \cite{de2021applying} and \cite{de2022reinforcement}). Since the assignment of integer values to $p$ as well as the additional time dependent constraints and variables would require too many qubits, we chose in our experiments a simplified version of the problem for the RL environment. A single episode consists of a sequence of \emph{contextual multiarmed bandit} problems: At the beginning of an episode we randomly sample 10 values for $L$ between $min(p_{min_i})$, the minimal amount of power the smallest generators can produce, and $\sum p_{max_i}$, the maximal amount of power all generators combined can produce. Then we also sample for each time step of the episode random values for all $p_i$ instead of assigning them by the agent. At each time step, the QRL agent outputs all values for the binary variables $x_i$. The reward at each time step equals the negative cost of objective function Eq. \ref{eq:uc_obj} for the chosen variables $x_i$. Finally, an episode ends after 10 time steps.

As for the weighted-MaxCut, we need to decide on a set of generators for our ansatz as well as a measurement strategy. The QUBO formulation of our problem exhibits both quadratic and linear terms, but unlike for the weighted-MaxCut, there is no need for annotations since we formulated our game as a sequence of \emph{contextual multiarmed bandit} problems. Nevertheless we introduce a trainable \emph{variational-block} consisting of a single generator $\sum^n_{i=1} X_i$, hence our generator set $G_{uc}$ for our QRL ansatz is 

\begin{equation} \label{eq:g_uc}
    G_{uc} = \Big\{\sum^n_{i<j} Z_iZ_j + \sum^n_{i=1} Z_i,  \sum^n_{i=1} X_i \Big\}.
\end{equation}

Since we need to assign all variables $x_i$ at each time step, we use single qubit observables $Z_i$ on all qubits. In future work, annotation strategies as well as additional constraints can be easily incorporated within our approach to tackle more realistic problem instances.

\paragraph{Knapsack}\label{sec:knapsack_env}

Let $N$ be the number of items, $v\in\mathbb{R}^N$ a vector that indicates the value of items $n=\{1,...,N\}$, $w\in\mathbb{R}^N$ a vector that indicates the weights of such items and $M\in\mathbb{R}$ the maximum weight. Then, the objective function of the KP can be formulated as

\begin{equation}
    max_{x\in\{0,1\}}\sum_{n=1}^{N}x_{i}v_{i}
\end{equation}

subject to the  inequality constraint

\begin{equation}
    \sum_{i=1}^{N}x_{i}w_{i}\leq M.
\end{equation}

The goal is to maximize the sum of the values of the chosen items while making sure that the total weight does not exceed the maximum weight. In our QRL approach, the inequality constraint will be encoded in the QUBO formulation using the \emph{unbalanced penalization} method, such that no additional qubits are required \cite{montanez2022unbalanced}. As has been stated in Section \ref{sec:inequality_encoding}, the ground state of the reformulated problem Hamiltonian might not be the same as for the original problem. Hence, the QRL agent is trained with the reward function in Eq. \ref{eq:reward}.

\begin{equation}\label{eq:reward}
    R = \begin{cases}
        x^Tv, & x^Tw \leq M\\
        0, & x^Tw > M
    \end{cases}
\end{equation}

Thus, even if the encoded problem Hamiltonian's ground state is not the optimal solution of the original CO problem, the model is trained on a ground truth reward function.

For Hamiltonian-based QRL, the generator set $G_{knapsack}$ is identical to $G_{uc}$ and we use single-qubit $Z_i$ observables on all qubits. The only difference is that the generator of the \emph{variational-block} is in fact an \emph{annotation-block} (ref. Fig. \ref{fig:qrl}). 

The RL environment of the KP is similar to the weighted-MaxCut: An episode starts with a randomly sampled problem instance from a dataset, with all annotation terms $\alpha$ set to $\pi$. At each time step, the QRL agent selects a single item and the selected item is then masked for all following time steps with its annotation value $\alpha$ set to 0. The reward function during training is Eq. \ref{eq:reward}. Since we need the complete vector $x$ (containing the values for all decision variables), the reward is only given in the last time step, with all other time steps yielding a reward of $0$. Moreover, an episode ends when the maximum weight is exceeded (let's assume at time step $t_f$), with the final vector $x$ being the vector at the end of time step $t_{f-1}$. The output of the QRL agent is the vector containing all the items before the constraint is broken, effectively masking invalid actions. Thus, the agent always outputs a valid action.

\section{Numerical Results}\label{sec:numerics}

In this section, we present the numerical results of Hamiltonian-based QRL on the three CO problems introduced, demonstrating that our method can be applied to solve all binary CO problems which have a QUBO formulation, much like QAOA. We thereby illustrate how different problem formulations and RL algorithms influence the performance of the Hamiltonian-based QRL agent. First, in Section \ref{maxcut} we start by using the weighted-MaxCut problem to compare our method to the one introduced in \cite{skolik2023equivariant}. Second, we show on the UCP how our ansatz outperforms the HEA ansatz as the number of qubits increases (see Section \ref{uc}). Third, the KP is used to benchmark our model against QAOA. The KP is used for the comparison with QAOA because it contains an inequality constraint, thus serving as an appropriate example to showcase all the capabilities of the introduced methods (see Section \ref{kp}). All models were trained on state vector simulators with Adam optimizers. For details on training parameters for all experiments as well as code to reproduce them we refer to GitHub \cite{github}.

\subsection{Weighted-MaxCut}\label{maxcut}

\begin{figure}[ht] 
\includegraphics[width=\linewidth]{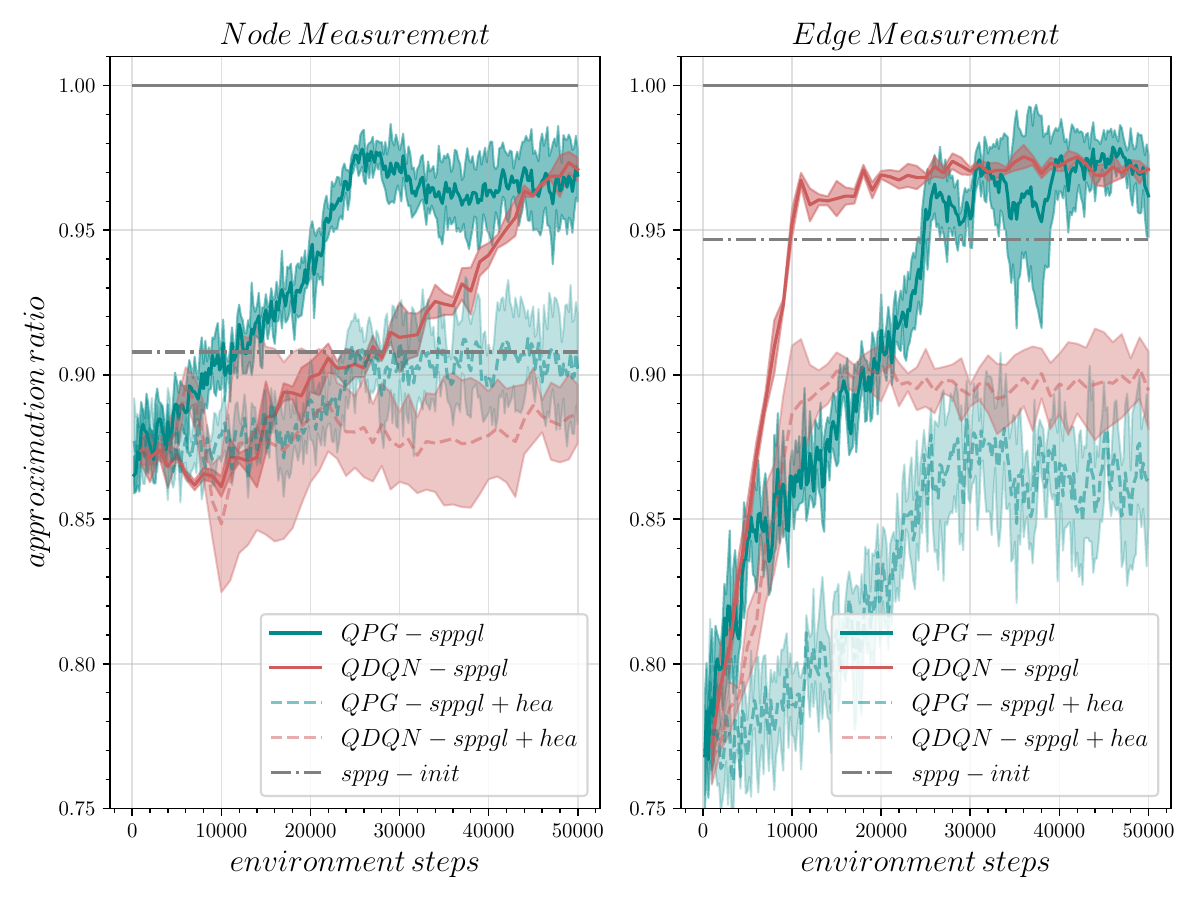}
\caption{Training performance of QDQN and QPG agents on weighted-MaxCut: Node and edge measurement strategy are compared for two RL algorithms with different ansatzes. The optimal value for the weighted-MaxCut environment is indicated by the solid gray line, while the average initial greedy policy of the QDQN with \textit{sge-sgv} ansatz is drawn as a dashed gray line. The main line of the QRL agents is the mean over 5 seeds with the shaded area indicating the standard deviation. }\label{fig:max_cut_exp}
\end{figure}

The weighted-MaxCut problem is the only CO problem considered in this work that can also be solved by the approach introduced by \cite{skolik2023equivariant}. Here, our ansatz only diverges slightly, since the Hamiltonian formulation of the problem only differs from the graph representation by a prefactor of $-\frac{1}{2}$. On this benchmark we compare the performance of two RL algorithms, QDQN and QPG, on two different versions of the weighted-MaxCut environment: We either measure the nodes or the edges during the action selection process. Additionally, we compare the performance of the \textit{sge-sgv} ansatz with the \textit{sge-sgv+hea} ansatz. The agents are trained for 50.000 time steps on a dataset of 100 graph instances with 5 nodes, hence VQCs with 5 qubits. We do not use additional output scaling for QDQN or QPG, but introduce a softmax temperature schedule for QPG, therefore preserving the equivariance property. Fig. \ref{fig:max_cut_exp} depicts the approximation ratio, which is defined as $r_{obtained}/r_{optimal}$, during training. However, we do not use this approximation ratio as a reward function (ref. Section \ref{sec:maxcut_env}), since this would require knowing the optimal solution beforehand.

Generally, the evaluated RL algorithms and action measurement strategies seem to converge to similar final scores when the \emph{sge-sgv} ansatz is used. However, the initial performance during the exploration phase of the algorithms has a different offset for the two measurement strategies. It can be clearly seen that the addition of the HEA ansatz hinders training, independently of the RL algorithm and action measurement strategy, illustrating the advantage of our ansatz.

An important feature of the \emph{sge-sgv} ansatz is that, for some graph-based problems like the weighted-MaxCut, the initialization of the ansatz already represents a competitive heuristic to solve the problem. We illustrate this feature in the following by analyzing the training behavior of the QDQN agent with \emph{sge-sgv} ansatz and edge measurement. However, all trained \emph{sge-sgv}-based agents exhibit this feature on weighted-MaxCut. The epsilon decay schedule from $1.0$ to $0.01$ is set for the first $10.000$ time steps. In the training plot it seems as if the agents improve their policy greatly during this initial phase. However, this is not the main explanation. Instead, the initial policy, indicated by the dashed black line, is already significantly better than the all-random policy. As epsilon decays, this shift towards the initial policy leads to the sharp increase, not the optimization of the policy. After the exploration phase is over, the increase in training performance is greatly decelerated. However, the agents still improve upon this initial policy. This means, that the main reason for the better performance of the \emph{sge-sgv} over the \emph{sge-sgv+hea} ansatz for the weighted-Maxcut is not mainly caused by the better training capabilities of the ansatz due to parameter correlation or symmetry-properties: Instead, the structure of the ansatz already represents a well performing heuristic.

\subsection{Unit Commitment}\label{uc}

\begin{figure*}[ht!] 
\centering
\includegraphics[width=\linewidth]{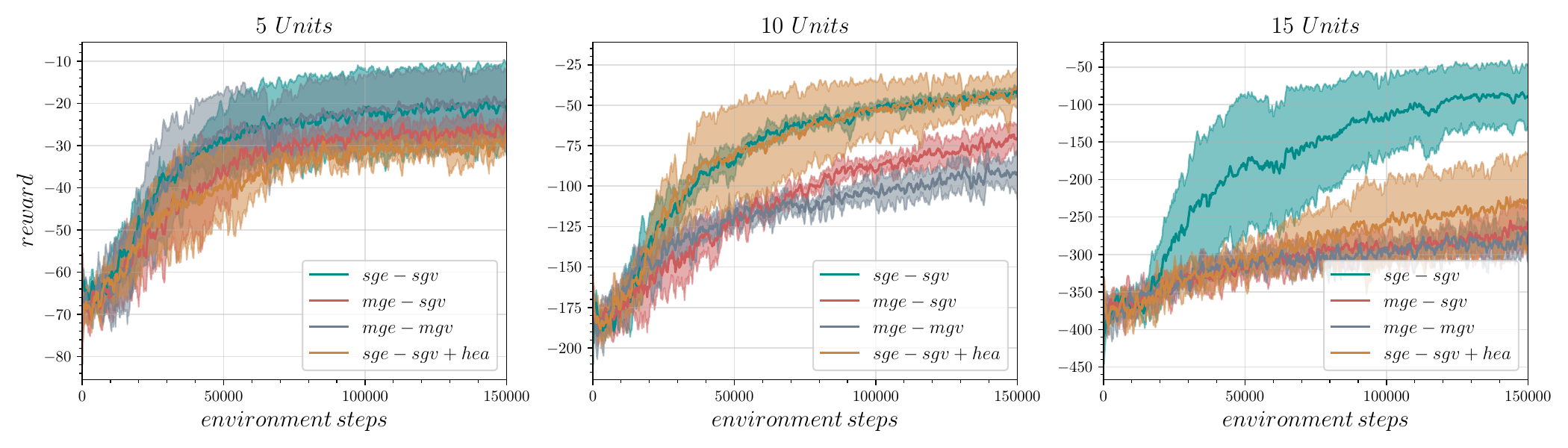}
\caption{Training performance of QPG agents on the UCP: Four ansatzes (\emph{sge-sgv}, \emph{mge-sgv}, \emph{mge-mgv}, \emph{sge-sgv+hea}) are evaluated on problem instances of size 5 (left), 10 (center) and 15 (right). The main line of the QRL agents is the mean over 5 seeds with the shaded area indicating the standard deviation.}\label{fig:uc_exp}
\end{figure*}

On the UCP we trained the QPG agent in a \emph{contextual multiarmed bandit} setting as described in Section \ref{sec:uc_env}. We evaluate all four ansatzes, which we introduced in Section \ref{H_based_QRL}, on this benchmark: the ansatzes \emph{sge-sgv}, \emph{mge-sgv} and \emph{mge-mgv} as well as the \emph{sge-sgv+hea} ansatz. We trained each agent for 150.000 time steps for problem instances of size 5, 10 and 15 generators and use VQCs with the same amount of qubits with $L=5$ layers. All ansatzes have additional trainable output scalings for all observables, hence the equivariance property is broken for all ansatzes. In  Fig. \ref{fig:uc_exp}, we plot the negative cost of energy production according to objective function Eq. \ref{eq:uc_obj} (which equals the reward) during the training of the QRL agents.

On the problem instances of size 5, all ansatzes show comparable training performance. Even though the \emph{sge-sgv} ansatz slightly outperforms the other ansatzes, this trend is not prominently indicated. On the larger problem instances of size 10 and 15 however, the advantage of the  \emph{sge-sgv} ansatz becomes clearly visible: even though the \emph{sge-sgv+hea} ansatz has still similar training performance for 10 units, only the \emph{sge-sgv} ansatz is able to successfully train on the 15 unit instances due to its greater trainability.

\subsection{Knapsack}\label{kp}

\begin{figure*}
        \centering
        \includegraphics[width=0.78\linewidth]{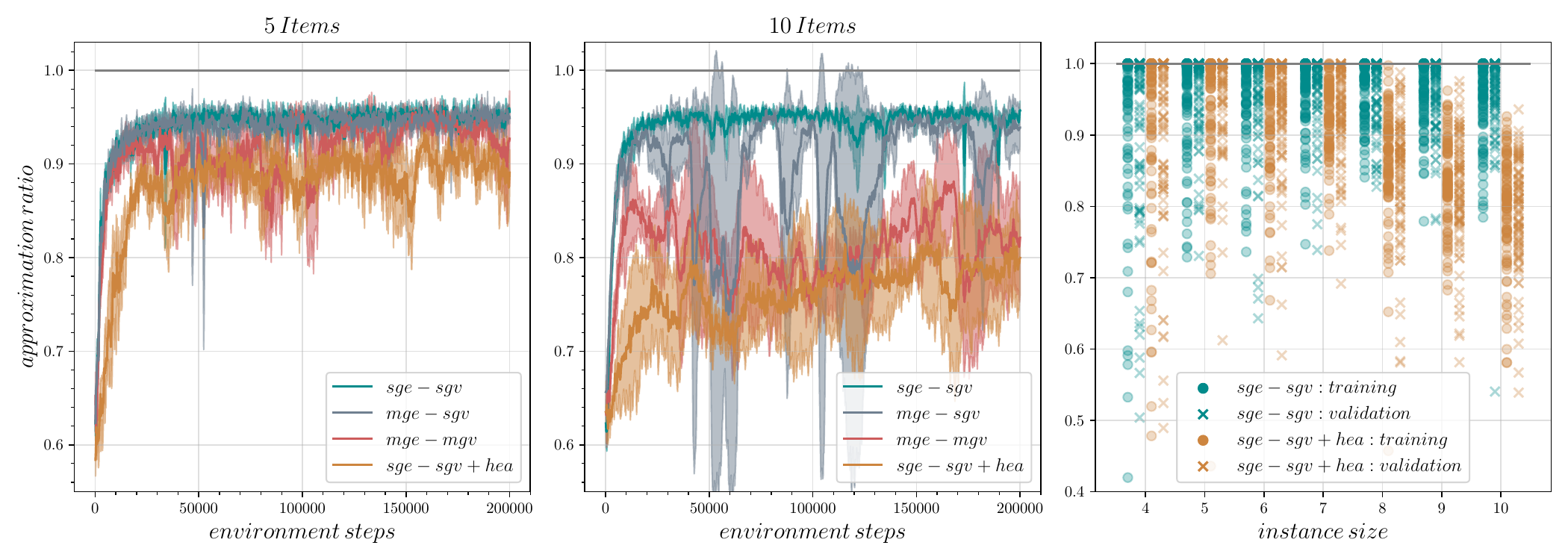}
        \caption{Training performance of QPG agents on KP: Four ansatzes (\emph{sge-sgv}, \emph{mge-sgv}, \emph{mge-mgv}, \emph{sge-sgv+hea}) are evaluated on problem instances of size 5 (left) and 10 (center). The main line of the QRL agents is the mean over 10 seeds with the shaded area indicating the standard deviation. The performance of \emph{sge-sgv} and \emph{sge-sgv+hea} ansatz are evaluated on 100 instances of the training dataset and 100 instances of the validation dataset for instance sizes 4 to 10 (right).}
    \label{fig:kp_main}
\end{figure*}

On the KP we trained QPG agents for each of the four ansatzes on an artificially generated dataset available in \cite{github} with $100$ Knapsack instances for $200.000$ time steps. In order to maintain the equivariance property of the VQC, the output scaling parameter needs to be the same across all qubits. Hence, we define a schedule that linearly increases this parameter throughout training. 

The first graph in Fig. \ref{fig:kp_main} shows the training performance on the 5 item instance of the KP. As for the UCP, all evaluated ansatzes are able to learn on this small problem instance. In the second graph of Fig. \ref{fig:kp_main} the training performance on the 10 item instance KP is depicted. While the \emph{mge-mgv} and the \emph{sge-sgv+hea} ansatz show poor training behaviour, the \emph{mge-sgv} ansatz still converges to a similar maximal score as the \emph{sge-sgv} ansatz, albeit being more unstable. In the third plot of Fig. \ref{fig:kp_main}, we compare the performance of the \emph{sge-sgv} and the \emph{sge-sgv+hwe} ansatz on the training dataset and an unseen validation dataset of $100$ Knapsack instances. As the instance sizes increase, the training as well as the validation performance of the \emph{sge-sgv} ansatz remains high, while the performance of the \emph{sge-sgv+hea} ansatz starts to decrease at instance sizes $>7$. Thus, similarly to what has been previously shown, our ansatzes consistently outperform HEA ansatzes at larger problem instances.

\begin{figure*}
    \centering
    \includegraphics[width=0.65\linewidth]{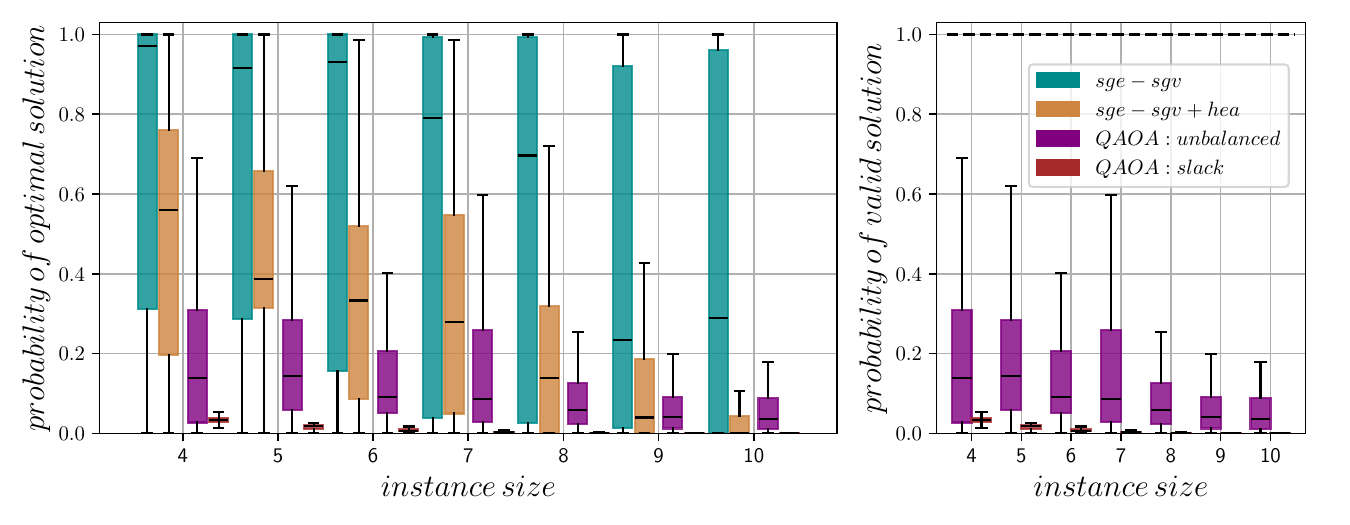}
    \caption{Probability of optimal action (left) and valid action (right) of QRL and QAOA on KP: The QRL ansatzes are evaluated on unseen KP problem instances of size 4 to 10 and benchmarked against QAOA models, which were trained using OpenQAOA with a maximum of 100 iterations and $p=3$ on each of the problem instances.}
    \label{fig:kp_vs_qaoa}
\end{figure*}

In Fig. \ref{fig:kp_vs_qaoa} we compare the final performance of the trained QRL agents with \emph{sge-sgv} and  \emph{sge-sgv+hea} ansatzes with QAOA. To do so, we trained $5$ QAOA models (using OpenQAOA with a maximum of 100 optimization steps and $p=3$ layers) with \emph{unbalanced} encoding and \emph{slack} encoding for each problem instance of the validation dataset (that contains $100$ instances) and averaged the probability of choosing the optimal or a valid action. The QRL agents were first trained on a training dataset with $100$ instances. For each model, $10$ agents were trained. Then, using the inference, for each of these agents we simulated $100$ episodes on each of the $100$ (unseen) problem instances of the validation dataset and averaged both the probability of choosing the optimal or a valid action over all agents and problem instances.

From the first graph of Fig. \ref{fig:kp_vs_qaoa}, it is clear that the QRL models find the optimal solution with a much higher probability than the QAOA models. Moreover, the \emph{sge-sgv} ansatz clearly outperforms the \emph{sge-sgv+hwe} ansatz, particularly as the instance size grows. From the second graph, one sees that the QAOA with \emph{unbalanced encoding} is capable of choosing a valid action (that does not break the constraint) with a much higher probability than the QAOA with \emph{slack} encoding. Moreover, as indicated  by the dashed line, the QRL agents always choose a valid action due to the hard constraint incorporated in their design.

\section{Discussion}
Neural Combinatorial Optimization (NCO) is a promising application area for Quantum Reinforcement Learning (QRL). The relation between QUBO formulation and problem Hamiltonian unveils a natural strategy to construct ansatzes for QRL. In this work, we introduced Hamiltonian-based QRL, where the ansatz is constructed from a small generator set inspired by the problem Hamiltonian of binary CO problems. The generator set consists of a single generator for the \emph{encoding-block} and a single generator for the \emph{variational-block}, hence \emph{sge-sgv} (\emph{single generator encoding-block - single generator variational-block}). We analyzed the trainability of our ansatz by calculating the variance of its gradient and found that it vanishes at a lower rate than the widely spread HEA or other variants of the \emph{sge-sgv} ansatz.

In order to demonstrate the board applicability and trainability of our approach, we applied it to a wide range of binary combinatorial optimization (CO) problems with sizes up to 15 qubits. On the weighted-MaxCut our \emph{sge-sgv} ansatz outperformed the HEA ansatz already at small qubit numbers. However, the main training advantage stems not from the improved trainability, but rather from a powerfull initial heuristic. Why the \emph{sge-sgv} ansatz encodes such a favourable heuristic for graph-based problems and how this property can be exploited for even better training performance will be left for future analysis. 

On the Unit Commitment problem, we analysed the training performance of our ansatz for instances from 5 to 15 qubits. While the advantage of the \emph{sge-sgv} ansatz over the other evaluated ansatzes was limited at small problem instances, it increased with problem size. For problem instances of size 15, only the \emph{sge-sgv} ansatz was able to stably train towards an optimal policy. These findings were confirmed on the Knapsack problem. On this benchmark we integrated the \emph{unbalanced penalization} method, originally proposed for QAOA \cite{montanez2022unbalanced}, in our QRL ansatz, such that the number of required qubits is reduced. By training our QRL agents with reward functions which depend on the original problem, we were able to cope with the disadvantage of the \emph{unbalanced penalization} method of potentially optimizing towards the ground state of an ill posed problem Hamiltonian.

The comparison of Hamiltonian-based QRL with QAOA showed that QRL consistently performs better, especially as problem instance size increases. QRL finds the optimal solution with a higher probability than QAOA at the cost of much higher required circuit evaluations during training. Moreover, QRL always finds a valid solution, since unfeasible solutions are masked during training, while QAOA often finds invalid solutions that break the constraints. Finally, QRL also generalizes well to unseen problem instances, which at least partially compensates for the increased training costs.

\section*{Acknowledgements}
The research is part of the Munich Quantum Valley, which is supported by the Bavarian state government with funds from the Hightech Agenda Bayern Plus.

\nocite{*}

\bibliographystyle{unsrt}
\bibliography{bibliography}

%




%
%

\end{document}